\documentclass[aps,prb,preprint,superscriptaddress, 12pt]{revtex4-2}

\usepackage[none]{hyphenat}
\usepackage{graphicx}
\usepackage{dcolumn}
\usepackage{bm}
\usepackage[utf8]{inputenc}
\usepackage[T1]{fontenc}
\usepackage{mathptmx}
\usepackage{textcomp}
\usepackage{siunitx}
\usepackage{bigints}
\usepackage{amsmath}
\usepackage{wrapfig}
\usepackage{ragged2e}
\usepackage{setspace}
\usepackage{hyperref}
\usepackage[switch]{lineno}
\usepackage[table,xcdraw,dvipsnames]{xcolor}
\usepackage{subcaption}
\usepackage{booktabs}
\usepackage{float}
\usepackage{color}
\usepackage{ulem}
\usepackage{multirow}
\usepackage[a4paper, left=2cm, right=2cm, top=3cm, bottom=3cm]{geometry}

\begin{document}
\singlespacing
\title{Elimination of substrate-induced FMR linewidth broadening in the epitaxial system YIG-GGG by microstructuring}

\author{David Schmoll}
\email{david.schmoll@univie.ac.at}
\affiliation{Faculty of Physics, University of Vienna, 1090 Vienna, Austria}
\affiliation{Vienna Doctoral School in Physics, University of Vienna, 1090 Vienna, Austria}
\author{Rostyslav O. Serha}
\affiliation{Faculty of Physics, University of Vienna, 1090 Vienna, Austria}
\affiliation{Vienna Doctoral School in Physics, University of Vienna, 1090 Vienna, Austria}
\author{Jaganandha Panda}
\affiliation{CEITEC BUT, Brno University of Technology, 61200 Brno, Czech Republic}
\author{Andrey A. Voronov}
\affiliation{Faculty of Physics, University of Vienna, 1090 Vienna, Austria}
\affiliation{Vienna Doctoral School in Physics, University of Vienna, 1090 Vienna, Austria}
\author{Carsten Dubs}
\affiliation{INNOVENT e.V. Technologieentwicklung, 07745 Jena, Germany}
\author{Michal Urbánek}
\affiliation{CEITEC BUT, Brno University of Technology, 61200 Brno, Czech Republic}
\author{Andrii V. Chumak}
\email{andrii.chumak@univie.ac.at}
\affiliation{Faculty of Physics, University of Vienna, 1090 Vienna, Austria}

\begin{abstract}

Modern quantum technologies and hybrid quantum systems offer the opportunity to utilize magnons on the level of single excitations. Long lifetimes, low decoherence rates, and a strong coupling rate to other subsystems propose the ferrimagnet yttrium iron garnet (YIG), grown on a gadolinium gallium garnet (GGG) substrate, as a suitable platform to host magnonic quantum states. However, the magnetic damping at cryogenic temperatures significantly increases due to the paramagnetic character and the highly inhomogeneous stray field of GGG, as recent experiments and simulations pointed out. Here, we report on temperature dependent ferromagnetic resonance (FMR) spectroscopy studies in YIG-GGG thin-films with different sample geometries. We experimentally demonstrate how to eliminate the asymmetric stray field-induced linewidth broadening via microstructuring of the YIG film. Additionally, our experiments reveal evidence of a non-Gilbert like behavior of the linewidth at cryogenic temperatures, independent of the inhomogeneous GGG stray field.   

\end{abstract}

\maketitle


\newpage

\section{Introduction}
The field of magnonics explores spin waves and their quanta, magnons, as the dynamic excitations of the collective spins in magnetically ordered materials. Continuous advances in nanofabrication technology and the growing need for innovation in data transportation and processing, focused attention to the potential of spin waves as data carriers in novel computing schemes~\cite{Chumak2022,Finocchio2024,Barman2021,Pirro2021,Mahmoud2020,Zeenba2024,Zeenba2024second,Wang2023,Heinz2020} and their implementation in wide frequency bandwidth RF devices~\cite{Levchenko2024,WangNature,Heussner2020,Vogt2014}. Alongside this classical area of spin-wave research, attention is also directed towards the quantum character of magnons. With their broad frequency range from GHz to THz, a set of experimentally accessible parameters to manipulate the dispersion characteristics, and the exhibited wavelengths, allowing circuit components down to a few nanometers, magnons emerge as a promising candidate for the integration in modern quantum technologies and in hybrid quantum systems~\cite{Chumak2022, Zhang2023, Jiang2023,Li2020,Quirion2019}. To utilize magnons on the level of single excitations, it is essential to maintain the coherence of the quantum state with low decoherence rates and achieve strong coupling with other quantum subsystems. Due to its uniquely low damping~\cite{Serga2010, LeCraw1958, CHEREPANOV199381}, the ferrimagnet yttrium iron garnet (YIG) became the material of choice for classical research in the field of magnonics, with various applications and devices already developed between 1960 to 1980~\cite{Adam1988, Glass1988, Ishak1988,Morgenthaler1988, Rodrigue1988}. The minimized dissipation, together with the large spin density~\cite{Cao2015}, propose YIG also as the suitable platform to investigate magnonic quantum states. 

To utilize magnons in future on-chip quantum technologies, experiments require millikelvin temperatures, to sufficiently suppress the population of thermally excited magnons and phonons, and high quality thin-film YIG samples, which are grown on gadolinium gallium garnet (GGG) substrates, due to the close matching of the lattice constants~\cite{Dubs2017}. However, at cryogenic temperatures, such YIG-GGG heterostructures face a significant increase in dissipation due to the paramagnetic character of GGG~\cite{Serha2024, Kosen2019}, expressed via a more than tenfold broadening of the ferromagnetic resonance (FMR) linewidth~\cite{Rosty2024}. This substrate-induced damping is also observed in the case of propagating magnons ($k>0$), with decreasing spin-wave transmission amplitudes at cryogenic temperatures~\cite{Schmoll2024, Knauer2023}. While the precise mechanisms and their contribution to the increase in magnon damping for uniform precessions are still not explored completely, micromagnetic simulations and experiments down to millikelvin temperatures addressed a parasitic linewidth broadening due the highly inhomogeneous magnetic stray field, created by the GGG substrate at decreasing temperatures~\cite{Serha2024, Rosty2024}.

Here, we report on FMR-spectroscopy measurements between room temperature and \SI{2}{\kelvin} in a thin-film YIG-GGG sample with different sample geometries, patterned via UV lithography and Argon ion beam etching. We show direct experimental evidence of the substrate-induced FMR-linewidth broadening and demonstrate how to eliminate its contribution to the magnon damping. Additionally, the experiments revealed a non-Gilbert like behavior of the FMR-linewidth at cryogenic temperatures and FMR-frequencies above approximately 18 GHz, independent of the parasitic stray field of the GGG substrate.

\section{Methodology}
In this work, we performed FMR-spectroscopy measurements for frequencies up to \SI{40}{\giga\hertz}, using a Rohde \& Schwarz model ZVA 40 vector network analyzer (VNA) and a commercially available broadband stripline antenna. The maximum applied microwave power was \SI{0}{\decibel}m. A Physical Property Measurement System (PPMS) enabled measurements in a temperature range from \SI{293}{\kelvin} to \SI{2}{\kelvin} with external bias fields $B_0$ up to \SI{1.3}{\tesla}, created by superconducting coils. The field was applied in-plane to the sample and parallel to the antenna. The experiments were conducted with YIG films of $4\times\SI{4}{\milli\meter}$ lateral size and \SI{150}{\nano\meter} thickness, grown by liquid phase epitaxy (LPE) in the $<111>$~-~crystallographic direction~\cite{Dubs2017}. In addition to the measurement of the FMR-signal at $B_0$, we performed reference measurements \SI{40}{\milli\tesla} above and below the target field. Subtraction of the averaged reference signal from the measurement at $B_0$ allows to minimize the magnetically active background signal of the partially magnetized GGG substrate at decreasing temperatures and investigate only the FMR-spectrum in the YIG layer. To allocate the FMR-frequency and fit the full linewidth at half maximum (FWHM), the background was analyzed with a 1D cubic spline model~\cite{herrera2023double} and the resonance shape was fitted with a split-Lorentzian model, individually describing the left and right side of the peak~\cite{Rosty2024}. As the FMR-signal was measured with a frequency sweep at a fixed external field $B_0$, the obtained linewidth $\Delta f$ was transformed to the magnetic linewidth $\Delta B$, taking into account the ellipticity of the angular spin-precession~\cite{Kalarickal2006}. At decreasing temperatures, the paramagnetic GGG substrate starts to magnetize and exhibits a magnetic stray field, which is antiparallel to the applied external field in the in-plane configuration~\cite{Serha2024}. This altering of the applied bias field $B_0$ and the increasing saturation magnetization of YIG at decreasing temperatures~\cite{CHEREPANOV199381}, was implemented in the calculation of $\Delta B$.

\begin{figure}
    \centering
    \includegraphics[width=1\linewidth]{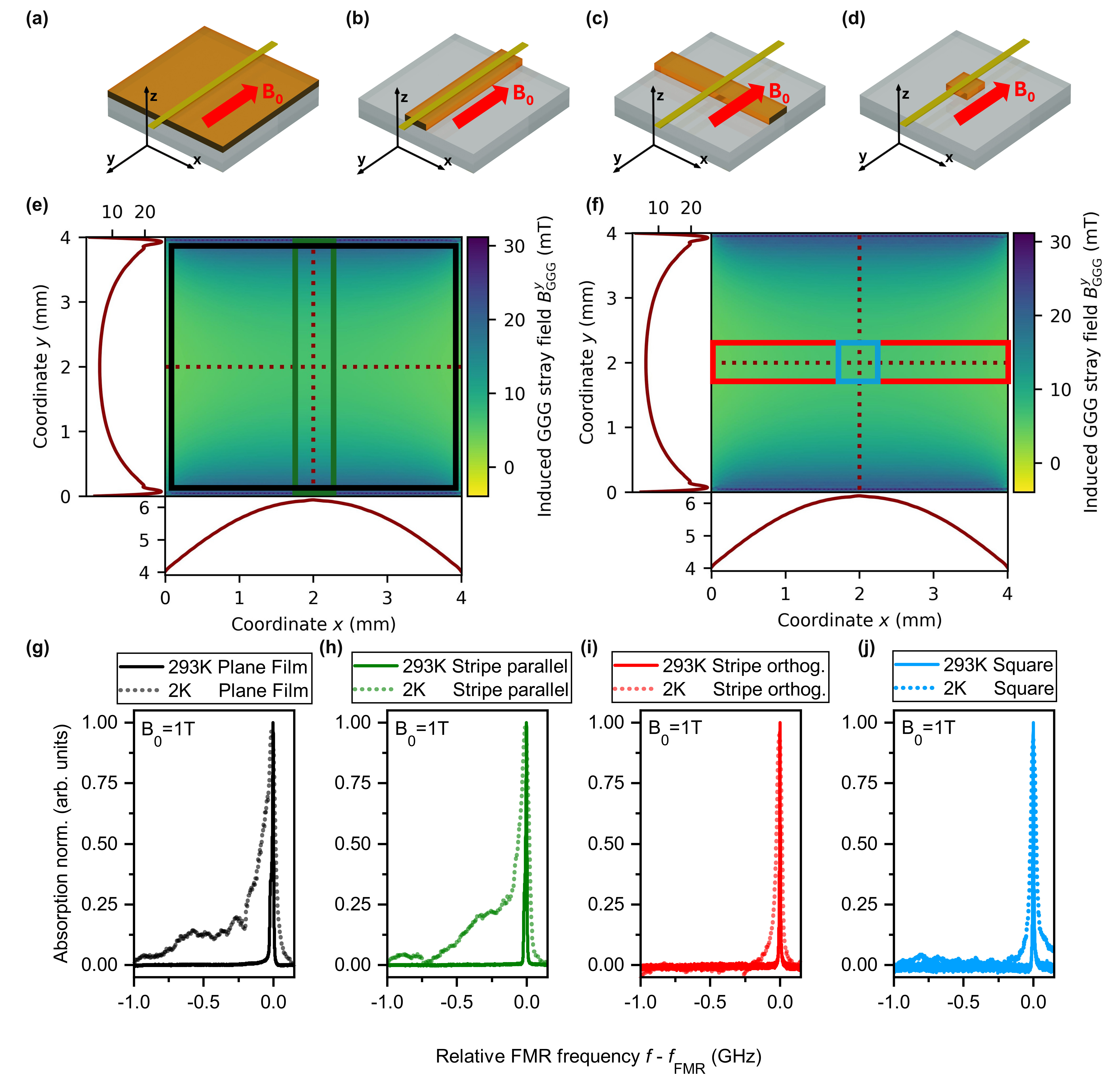}
    \captionsetup{justification=justified}
    \caption{\justifying (a)-(d) Different experimental configurations of the temperature dependent FMR-measurements. The \SI{150}{\nano\meter}-thick YIG films are depicted in brown, the GGG substrate in grey, and the stripline antenna is indicated in gold. Measurements were performed with a plane YIG film ($4\times$\SI{4}{\milli\meter}), a microstructured YIG stripe ($4\times$\SI{0.5}{\milli\meter}) placed parallel to the external field $B_0$, a microstructured YIG stripe placed orthogonal to the external field $B_0$, and a microstructured YIG square ($0.5\times$\SI{0.5}{\milli\meter}). (e)-(f) Magnitude of the GGG-induced stray field $B_\mathrm{GGG}$ (colormap) and projection of the field magnitude along the lateral coordinates of the YIG film. The sample positions in the field are indicated as colored boxes. The field was simulated numerically with the Maxwell solver FEMME. (g)-(j) FMR-spectra recorded in the PPMS at room temperature (solid lines) and at \SI{2}{\kelvin} (dotted lines) with an external bias field $B_0=$ \SI{1}{\tesla}. The following FMR-frequencies were observed for the depicted spectra: (g) $f_\mathrm{FMR}^\mathrm{293K}=$ \SI{30.60}{\giga\hertz}, $f_\mathrm{FMR}^\mathrm{2K}=$ \SI{30.90}{\giga\hertz}, (h) $f_\mathrm{FMR}^\mathrm{293K}=$ \SI{30.60}{\giga\hertz}, $f_\mathrm{FMR}^\mathrm{2K}=$ \SI{30.94}{\giga\hertz}, (i) $f_\mathrm{FMR}^\mathrm{293K}=$ \SI{30.56}{\giga\hertz}, $f_\mathrm{FMR}^\mathrm{2K}=$ \SI{30.83}{\giga\hertz}, (j) $f_\mathrm{FMR}^\mathrm{293K}=$ \SI{30.56}{\giga\hertz}, $f_\mathrm{FMR}^\mathrm{2K}=$ \SI{30.80}{\giga\hertz}.}
    \label{Fig1}
\end{figure}

To investigate the linewidth broadening, induced by the inhomogeneous stray field of the GGG substrate, we compared a plane YIG film with two samples from the same wafer, but with different geometries patterned into the film via UV direct-write lithography and consequent Argon ion beam etching. We used \SI{1.5}{\micro\meter} thick negative resist AZ nLOF 2020 as the etch mask for a \SI{600}{e\volt} broad Ar$^+$ ion beam. On the first fabricated sample, YIG was removed down to GGG, except for a $0.5\times\SI{0.5}{\milli\meter}$ square region in the middle of the film. For the second sample, YIG was removed down to the GGG surface, except for a $0.5\times\SI{4}{\milli\meter}$ stripe across the center of the film. Figures~\ref{Fig1}~(a)-(d) depict the four different measurement configurations in which the FMR signal was recorded: (a) the plane YIG film, (b) the YIG stripe magnetized along its length and placed parallel to $B_0$, (c) the YIG stripe magnetized along its width and placed orthogonal to $B_0$, and (d) the YIG square. In this way we can compare the FMR-signal and the linewidth obtained from the fitting model for the sample exposed to the inhomogeneity of the stray field - configuration Fig.~\ref{Fig1}~(a) and (b) - and for the sample exposed only to the homogeneous area of the substrates stray field - configuration Fig.~\ref{Fig1}~(c) and (d). To visualize and quantify the substrates stray field inhomogeneity, we performed numerical simulations, using the FEMME software, solving the nonlinear Maxwell equations for the temperature and field-dependent magnetization of GGG, determined experimentally via vibrating sample magnetometry (see~\cite{Serha2024}). The calculated magnetic stray field was evaluated \SI{75}{\nano\meter} above the GGG surface in the area of the YIG film and is illustrated in Fig.~\ref{Fig1}~(e) and (f).
\section{Results}

The different sample geometries, measured in the configurations sketched in Fig.~\ref{Fig1}~(a)~-~(d), allow to visualize the asymmetric broadening of the recorded FMR-peaks and provide direct experimental evidence, that it originates from the inhomogeneity of the GGG stray field. In the case of the plane YIG film and the microstructured YIG stripe, placed parallel to the stripline antenna, the specimen is exposed to the inhomogeneous area of the substrates stray field (see Fig.~\ref{Fig1}~(e)). As the stray field exhibits a higher field magnitude $B_{\mathrm{GGG}}$ at the edges of the sample, especially along the y-direction, and the field is directed antiparallel to the target field in the case of in-plane magnetization~\cite{Serha2024}, the FMR-resonance condition is shifted towards lower frequencies for these sample regions, leading to the observed asymmetric broadening. This parasitic effect can be circumvented by placing the sample only in the homogeneous region of the substrates stray field, which was experimentally realized by placing the YIG stripe orthogonal to the external field $B_0$ and by the microstructured square YIG sample (see Fig.~\ref{Fig1}~(f)). 
Figures~\ref{Fig1}~(g)~-~(j) depict the FMR-spectra obtained in all four measurement configurations at room temperature (solid lines) and at \SI{2}{\kelvin} (dotted lines) for an external bias field $B_0=$ \SI{1}{\tesla}. The FMR-signal is plotted as the absorption $S_{21}/S_{21,\mathrm{ref}}$ normalized for the maximum value, versus the relative shift to the FMR-frequency $f-f_{\mathrm{FMR}}$. The direct comparison of the spectra recorded at room temperature and at \SI{2}{\kelvin} highlights the asymmetric broadening towards lower frequencies, due the substrates stray field, for the plane YIG film (black) and the YIG stripe placed parallel to the external field (green). The measurements of the YIG stripe placed orthogonal to $B_0$ (red) and the microstructured YIG square (blue) demonstrate, that we can avoid this substrate-induced broadening, when the investigated YIG film is only exposed to the homogeneous area of the GGG stray field. 
\begin{figure}[b]
    \centering
    \includegraphics[width=1\linewidth]{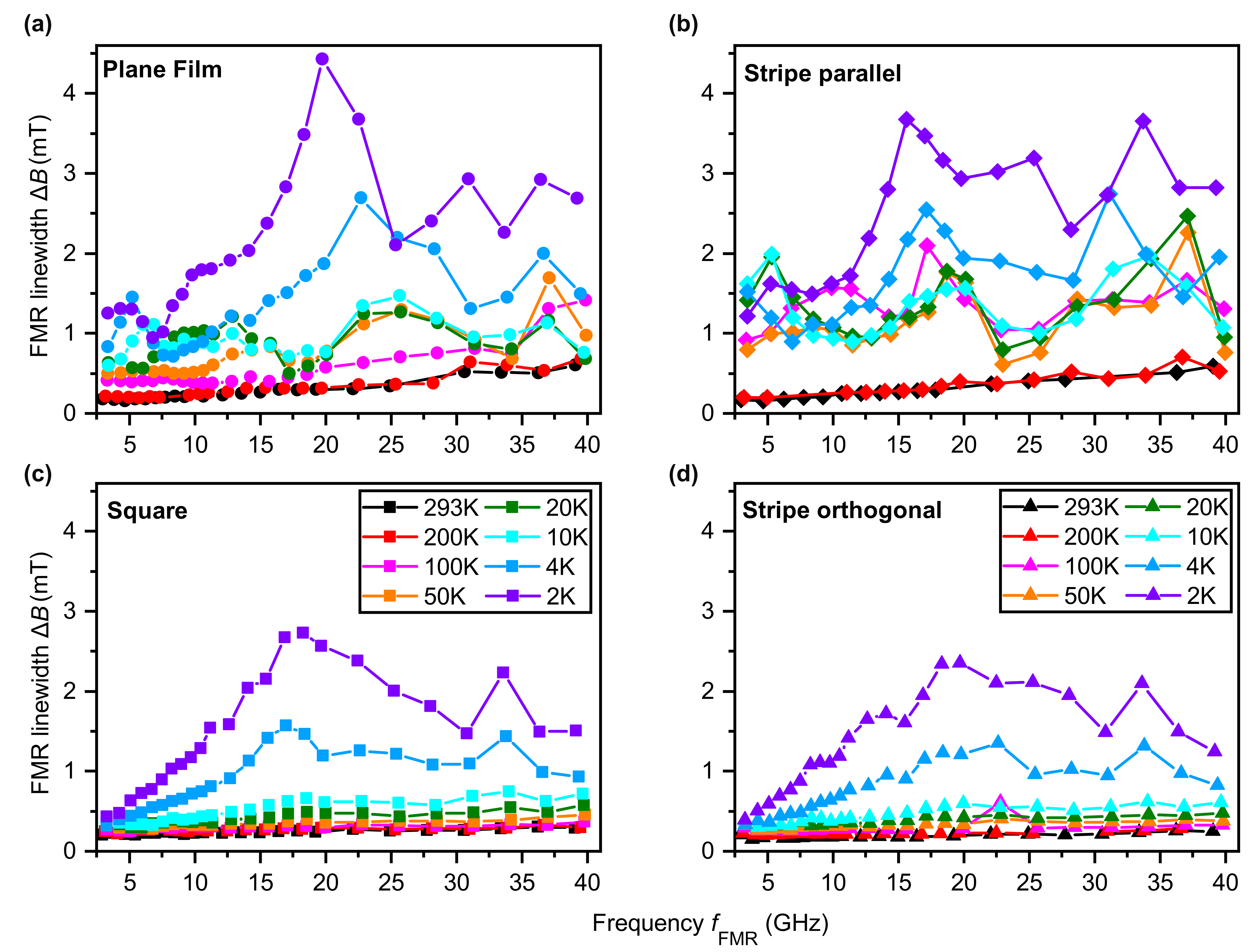}
    \captionsetup{justification=justified}
    \caption{\justifying Linewidth $\Delta B$, extracted from temperature dependent FMR-measurements between \SI{293}{\kelvin}~-~\SI{2}{\kelvin} and depicted versus the FMR-frequency $f_\mathrm{FMR}$. The different measurement temperatures are illustrated by different colors indicated in the legends of panel (c) and (d). The experiments were performed for external fields $B_0$ up to \SI{1.3}{\tesla} in four different measurement configurations: (a)-(b) the YIG film is exposed to the inhomogeneous area of the GGG stray field. (c)-(d) the YIG film is only exposed to the homogeneous area of the GGG stray field.}
    \label{Fig2}
\end{figure}
In this case, we observe no asymmetric broadening of the absorption spectrum. Our measurements emphasize, that the position of the investigated film with respect to the substrates edges and the magnetization direction, indeed plays a critical role in experiments with YIG-GGG samples at cryogenic temperatures. 

Due to the split-Lorentzian fitting model, we can extract the linewidth $\Delta B$ from all recorded spectra, even with the observed asymmetric broadening. We recorded the FMR-signal for target fields ranging from $B_0=$ \SI{50}{\milli\tesla} up to $B_0=$ \SI{1.3}{\tesla} at various temperatures between \SI{293}{\kelvin} and \SI{2}{\kelvin}. The applied fields correspond to a measurement frequency range of \SI{3}{\giga\hertz}~-~\SI{40}{\giga\hertz}, with the recorded FMR-frequencies depending on the temperature, as they experience a shift due to the increasing saturation magnetization of YIG and the stray field of the partially magnetized GGG substrate. Figures~\ref{Fig2}~(a)-(d) depict the extracted linewidth $\Delta B$ versus the FMR-frequency $f_\mathrm{FMR}$, recorded for all four measurement configurations. The measurement temperatures are indicated by different color coding. When the sample is exposed to the inhomogeneous area of the GGG stray field (see Fig.~\ref{Fig2}~(a) and (b)), we observe a linear increase of $\Delta B$ across the complete measured frequency range for temperatures down to \SI{200}{\kelvin}. Below \SI{200}{\kelvin}, a general increase of the linewidth $\Delta B$ with a decrease in temperature is recorded. However, for temperatures between \SI{200}{\kelvin} and \SI{10}{\kelvin}, we observe an oscillation of the linewidth with frequency, instead of the expected linear increase. This behavior was also reported in recent experiments, studying the increase of effective magnetic damping in YIG-GGG thin-films down to \SI{30}{\milli\kelvin}~\cite{Rosty2024}. We partly attribute these oscillations to inhomogeneities in the thickness of the YIG film, which lead to temperature and magnetic field-dependent strain effects along the sample. Below \SI{10}{\kelvin}, the linewidth increases non-linear with frequency to a temperature dependent maximum value at approximately \SI{18}{\giga\hertz}. After the maximum linewidth, we again observe oscillations of $\Delta B$ with an increase in frequency.

\begin{figure}[t]
    \centering
    \includegraphics[width=1\linewidth]{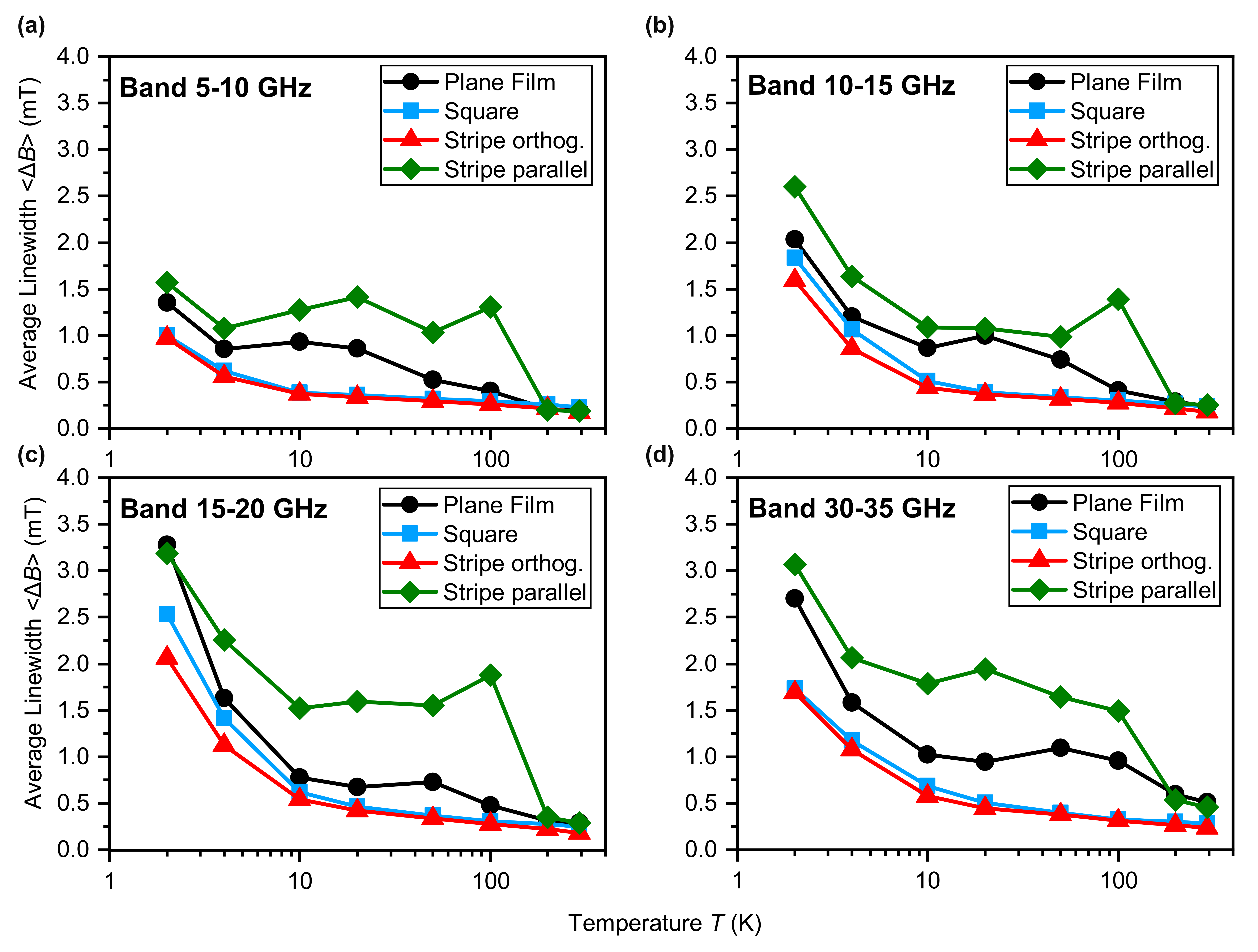}
    \captionsetup{justification=justified}
    \caption{\justifying Averaged linewidth $<\Delta B>$ versus temperature. The data were averaged for the indicated frequency intervals, to allow consistent comparison between the four different measurement configurations.}
    \label{Fig3}
\end{figure}
When the YIG film is only exposed to the homogeneous region of the substrate-induced stray field (see Fig.~\ref{Fig2}~(c) and (d)), we observe a linear increase of $\Delta B$ with frequency for temperatures down to \SI{50}{\kelvin}. Between \SI{50}{\kelvin} and \SI{10}{\kelvin}, the measurements also reveal small oscillations of the linewidth with an increase in frequency. But as the FMR-signal is detected without the parasitic influence of the stray field inhomogeneity and not along the complete film but only in a small area placed on the antenna, the oscillations are strongly suppressed compared to Fig.~\ref{Fig2}~(a) and~(b). However, a decrease of temperature below \SI{10}{\kelvin} again shows a non-linear increase of the linewidth to a maximum value at approximately \SI{18}{\giga\hertz}, suggesting that this non-Gilbert behavior is not caused by the inhomogeneous GGG stray field. Due to the suppressed oscillations, we can observe a steady decrease of the linewidth with increasing frequencies, after the maximum of $\Delta B$ is reached. This trend underpins, that the magnetic damping at temperatures below \SI{10}{\kelvin} does not follow the classic Gilbert model anymore, especially at high external magnetic fields $B_0$. Although the physical mechanisms of the experimentally observed linewidth behavior are not yet clear, the temperature dependence indicates a connection to the magnetization of the GGG substrate. 

To illustrate the observed increase of the linewidth with temperature for the different measurement configurations, the extracted values of $\Delta B$ were averaged for certain frequency bands. Figures~\ref{Fig3}~(a)-(d) depict the averaged linewidth $<\Delta B>$ for the frequency intervals 5~-~\SI{10}{\giga\hertz}, 10~-~\SI{15}{\giga\hertz}, 15~-~\SI{20}{\giga\hertz}, and 30~-~\SI{35}{\giga\hertz}. The general increase of $<\Delta B>$ with decreasing temperatures can be observed for all measurement configurations and across all frequency intervals, indicating higher effective magnetic damping at cryogenic temperatures. When the YIG film is exposed to the inhomogeneous area of the GGG stray field, we observe a linear increase of $<\Delta B>$ with temperature down to \SI{200}{\kelvin} across all frequency intervals, followed by an oscillation behavior down to \SI{4}{\kelvin} for frequencies below \SI{10}{\giga\hertz} and down to \SI{10}{\kelvin} for frequencies above \SI{10}{\giga\hertz}. If the sample is exposed only to the homogeneous area of the GGG stray field, the averaged linewidth $<\Delta B>$ increases linearly with decreasing the temperature down to \SI{10}{\kelvin} for all frequency intervals. Below \SI{10}{\kelvin}, we observe a non-linear increase of the linewidth with decreasing temperatures, independent of the measurement configuration. These observed trends are in agreement with Fig.~\ref{Fig2}~(a)-(d). Although the linewidth generally is smaller without the influence of the inhomogeneous stray field, the non-linear increase of $<\Delta B>$ below \SI{10}{\kelvin} is visible in all measurement configurations and therefore of different nature. The understanding of the physical mechanism behind the non-linear behavior of the linewidth and the contributions from the partially magnetized GGG substrate requires further theoretical considerations.

\section{Conclusion}
We demonstrate how to eliminate the asymmetric broadening of the FMR-signal in YIG thin-films grown on GGG, introduced by the inhomogeneous stray field, which originates from the partially magnetized substrate at decreasing temperatures. By microstructuring the plane YIG film via UV lithography and Argon ion beam etching to a YIG stripe and a YIG square, small enough to fit in the homogeneous area of the GGG stray field, we eliminated the substrate-induced asymmetric broadening. Additionally, FMR measurements for frequencies up to \SI{40}{\giga\hertz} and temperatures between \SI{293}{\kelvin} and \SI{2}{\kelvin} revealed a deviation from the expected Gilbert-like linear increase of $\Delta B$ with frequency at cryogenic temperatures and instead exhibits oscillations of the linewidth magnitude with frequency. Our experiments illustrate that the oscillations are still present but can be strongly suppressed, if the YIG specimen is exposed only to the homogeneous area of the GGG stray field and the FMR-signal is not detected along the whole film but only in a small sample area on the antenna. Below temperatures of \SI{10}{\kelvin}, we observed a non-linear linewidth increase with a temperature dependent maximum value approximately at \SI{18}{\giga\hertz}, followed by a non-Gilbert behavior where the linewidth decreases with increasing frequency. This behavior was recorded for the YIG plane film and for the microstructured samples, with no exposure of the YIG film to the inhomogeneous GGG stray field, suggesting that the physical reason behind the non-linear nature is not related to the substrate-induced magnetic field. Our results emphasize the parasitic nature of the substrate-induced magnetic stray field and demonstrate how to avoid its effect in future YIG-GGG based on-chip quantum circuits. 

\section*{Acknowledgments}
This work has been supported by the Austrian Science fund FWF in the frame of the project Paramagnonics (10.55776/I6568). The presented computational results have been achieved using the Vienna Scientific Cluster (VSC). C.D. gratefully acknowledges the support of the Deutsche Forschungsgemeinschaft (DFG, German Research Foundation) under Grant No. 271741898. M.U. acknowledges the support from project No. CZ.02.01.01/00/22008/0004594 (TERAFIT). Czech NanoLab project LM2023051 is acknowledged for the financial support of the sample fabrication at CEITEC Nano Research Infrastructure. We would also like to thank Roman Verba and Sebastian Knauer for valuable discussions.

%

\end{document}